\documentclass[a4paper,10pt]{article}
\usepackage{amsfonts, amsthm, latexsym}
\usepackage{amssymb,amsmath, graphicx}
\usepackage{dsfont}
\usepackage{subfigure}

\newtheorem{lemma}{Lemma}[section]
\title{IMEX schemes for a Parabolic-ODE system of European Options with Liquidity Shocks}
\author{ W.\; Mudzimbabwe,\;\;Lubin\;G.\;Vulkov}
\date{{\it Department of Applied Mathematics, Ruse University, Studentska str. 8, 7017 Ruse}}
\begin{document}
\maketitle

\begin{abstract}
The coupled system, where one is a degenerate parabolic equation and the other has not a diffusion term arises in the modeling
of European options with liquidity  shocks.
Two implicit-explicit (IMEX) schemes
that preserve the positivity of the differential problem solution
are constructed and analyzed.
Numerical experiments confirm the theoretical results and illustrate the
high accuracy and efficiency of the schemes in combination with Richardson extrapolation
\end{abstract}


\section{Introduction}

We study numerically a parabolic-ODE system modelling option pricing liquidity shocks . The presence of liquidity shocks is a source of
non-liquidity risk and makes this market incomplete.
Ludkowsky and Shen [5] investigate a nonlinear pricing mechanism based on utility
maximization.
They consider the investor whose utility is described by an exponential utility function

$$
\mathbf{\mathcal{U}} (x)=-e^{-\gamma x},
\eqno{(1)}
$$

\noindent where $\gamma>0$ is the coefficient of risk aversion. The investor seeks to maximise utility of both terminal wealth and option payoff at
 time horisont $T < \infty$, which is chosen to coincide with the expiration date of all securities in market model.
%
%
%
%
Properties of the exponential utility function (1) imply  that the value functions   can be expressed as

$$ {\widehat{   U}}^{i} (t,X,S) =
-e^{ - \gamma X } e^{ - \gamma R^{i} (t,S)  }, i=0,1,
\eqno{(2)}
$$

\noindent where
$X=X_{t}$ is the wealth process and the functions
$R^{i} (t,S)$
are related to the price of options in the two states, see (7) below.
Then the pair $\{ R^{i} (t,S), \; i=0,1 \}$ is the unique viscosity solutions of the coupled semi-linear system,

\setcounter{equation}{2}

\begin{align}
\label{smallpde}
\left.
\begin{array}{l}
R_{t}^{0} +\frac{1}{2}\sigma^{2} S^{2}R_{SS}^{0} -\frac{  \nu_{01} }{  \gamma }
e^{ - \gamma  (R^{1}- R^{0} ) }+\frac{ (d_{0} + \nu_{01} ) }{\gamma } =0,\\
R_{t}^{1} -
\frac{  \nu_{10}  }
{\gamma }
e^{ - \gamma  (R^{0}- R^{1} ) }
+
\frac{  \nu_{10}  }
{\gamma }=
0 .
\end{array}
\right.
\end{align}

\noindent The terminal conditions are:

$$
R^{i} (T,S) = h(S) , \; \; i=0,1. \eqno{(4)}
$$
 Here $\sigma$ is volatility of the underlying, $\nu_{01}, \nu_{10}$ are transition intensities from state (0) to state (1) and vice versa, respectively, $\mu$ is drift of the underlying and $d_0 = \mu^2/2\sigma^2$, see [5] for more details.

Using $ {\widehat{U}}^{i} $ and $ {\widehat{V}}^{i} $ , the buyer's indifference price $p$ (initial state 0) and $q$ (initial state
1) are defined via

$$
{\widehat{U}}^{0} (t,X - p,S) =
{\widehat{V}}^{0} (t,X) , \; \;
{\widehat{U}}^{1} (t,X - q,S) =
{\widehat{V}}^{1} (t,X),
\eqno{(5)}
$$

\noindent where $ {\widehat{U}} ,
{\widehat{V}} $ are the optimal solutions for terminal wealth with and without options respectively.
The value functions ${\widehat{V}}^{i}, \; i=0,1$ are given by

$$ {\widehat{V}}^{i} = e^{ - \gamma X}  F_{i} (t) \; \; \mbox{ and} \; \;
 {\widehat{V}}^{i} (t, X, S) = e^{  - \gamma R^{i} (t,S)  } , \; \; i=1,2
 \eqno{(6)}
 $$


\noindent and the functions $F_{0} (t), F_{1}(t)$  by, \begin{align}
F_{0} (t) &=c_{1} e^{  \lambda_{1} t } + c_{2} e^{  \lambda_{2} t }, \nonumber\\
F_{1} (t) &= \dfrac{1}{\nu_{01} }\{c_{1}(d_{0} +\nu_{01} - \lambda_{1} )e^{  \lambda_{1} t } + c_{2}(d_{0} +\nu_{01} - \lambda_{2} )e^{  \lambda_{2} t } , \nonumber
\end{align}
 where \begin{align}
&  \lambda_{1,2} =\dfrac{d_{0} + \nu_{01} +\nu_{10} \pm {\sqrt{ (d_{0} +\nu_{01} +\nu_{10})^{2} - 4 d_{0}\nu_{10}
}}}{2} ,\nonumber\\
& c_{1}
= \dfrac{\lambda_{2} - d_{0}}{\lambda_{2} - \lambda_{1}}e^{  - \lambda_{1} T }\; \;\mbox{and} \; \;c_{2} =\dfrac{\lambda_{1} - d_{0}}{\lambda_{1} - \lambda_{2}}e^{  - \lambda_{2} T }.\nonumber
\end{align}

\noindent Then, we obtain from (2), (5), (6)

$$
p=R^{0} + \gamma^{-1} \ln F_{0} (t), \; \;
q=R^{1} + \gamma^{-1}   \ln F_{1} (t)
\eqno{(7)}
$$

\noindent and from (3), (4) the  parabolic-ordinary system  for $p$ and $q$


$$
\begin{array}{l}
 p_{t}  +\frac{1}{2}\sigma^{2} S^{2} p_{SS} -\frac{  v_{01} }{  \gamma }
 \frac{ F_{1} }{ F_{0}   }
e^{ - \gamma  ( q- p ) }+\frac{ (d_{0} + v_{01} ) }{\gamma }
- \frac{1}{\gamma } \frac{ F_{0}' }{ F_{0}   } = 0,
\\
 q_{t}
  -\frac{  v_{10} }{  \gamma }
 \frac{ F_{0} }{ F_{1}   }
e^{ - \gamma  ( q- p ) }+\frac{ v_{10}  }{\gamma }
- \frac{1}{\gamma } \frac{ F_{1}' }{ F_{1}   } = 0
\end{array}
\eqno{(8)}
$$

\noindent with terminal conditions

$$
p(T,S) = q (T,S) = h (S).
\eqno{(9)}
$$




The numerical solution of the system (8) is the main object of the present paper. The numerical treatment of the boundary layer effect for small values of $\sigma_{0} $ and $\gamma $ , the degeneracy at $S=0$ of the parabolic equation and the exponential nonlinearity lead to challenging  problems [10].
The introduction of exponential nonlinear terms is an available assumption based on the financial nature of the model system (8).
There are many numerical schemes to solve nonlinear parabolic and hyperbolic equations. However, very few have dealt with an exponential
nonlinear term. The special nature of the nonlinear exponential term for a hyperbolic problem is discussed in [10].
A possible way to build an efficient numerical solution of (8), (9) is to implement an
IMEX method [1,9] . {\it{In this procedure}} the diffusion term is discretized implicitly in time and the reaction terms are discretized explicitly.

An IMEX method for numerical solution of reaction-diffusion equation with pure Neumann boundary conditions is developed in [3].
IMEX schemes, by applying explicit approximation both integral term and the convection term and an implicit approximation for
the second differential term are developed for integro-differential equations of finance in [2].

The rest of the paper is organized as follows. In the next section some results concerning well-posedness of Cauchy problem for (\ref{smallpde})
and a comparison principle, obtained in [4] , are discussed . Also, two lemmas  concerning discrete maximum principle [6,7] are formulated.
In Section 3 an implicit-explicit linear scheme is introduced. Comparison discrete principle and convergence of the scheme are
proved. Similar results are obtained for an IMEX
linearized scheme in Section 4. The computational experiments
in Section 5
confirm the applicability of our schemes and the theoretical results.
Finally, Section 6 summarizes our conclusions.

{\bf{   Notation }}
Let $\Omega $ be a bounded interval in $R^{+} = ( 0 , \infty)$ and let
$C^{0} ( \Omega ) $ denote the space of continuous functions on $\Omega $  with the norm of any $ w \in C^{0} ( \Omega ) $ defined by
$ \Vert w \Vert_{ \Omega  } =
\sup_{  x \in \Omega } \vert w(x) \vert$. For each
integer   $ k \geq 1 $ , let  $ C^{k} ( \Omega )$ denote the space of $k$ - times differentiable functions on $ \Omega $, with
continuous derivatives up to and including those of order $k$, with the norm of any  $ w \in C^{k} ( \Omega )$ defined by
$ \Vert w \Vert_{ k, \Omega  } =
\max_{  0 \leq l \leq k } \Vert w(x)^{(l)} \Vert_{ \Omega }$.
The notational conventions $ \vert w \vert_{ 0, \Omega  } = \Vert w \Vert_{ 0, \Omega  } = \Vert w \Vert_{ \Omega  } =
\Vert w \Vert $ are adopted . The explicit reference to $\Omega $ is dropped whenever the domain question is evident.
For any mesh functions on arbitrary mesh
$ \Omega^{N} = \{  x_{i} \}_{1}^{N-1} $,
$ {\overline { \Omega   }}^{N} =
\{  x_{i} \}_{0}^{N} $ the discrete maximum norm is defined by
$ \Vert w \Vert_{ C( {\overline{ \Omega} ) }^{N}  } = \max_{   0 \leq i \leq N } \vert w_{i} \vert  .$

Maximum norms and semi-norms for smooth functions of two variables are introduced in a similar way. Let
$Q_{T} =(0,T)  \times \Omega   $. Then
$ \Vert w \Vert_{ Q_{T}   } =
\sup_{  (x, \tau) \in Q_{T} } \vert w(x, \tau) \vert$ and if $C^{0} ( Q_{T} )$ is the space of all functions on
$Q_{T} $  with continuous derivatives then

$$
C^{k} ( Q_{T} ) =
\left\{ w : \frac{  \partial^{i+j} w }
 {
 \partial x^{i} \partial \tau^{j}
  } \in C^{0} ( Q_{T} ) \; \; \mbox{ for }
  \; \;
  i,j=0,1,2, \dots \; \; \mbox{ with } \; \;
  0 \leq i + 2 j \leq k \right\} . $$

\section{Preliminaries }
In this section we will describe some properties of the solution to system (8) using results obtained in [4].
Also
, following  [6,7],
 two lemmas , concerning discrete maximum principle (DM) are formulated.


We will consider solutions of (8)
 satisfying

$$
\left|p\right|,\left|q\right|,\left|h\right|\leq A\exp\left(\alpha\ln^2S\right)=AS^{\alpha\ln S},
\eqno{(10)}
$$
\noindent for some positive constants $A$ and $\alpha$.
In [4], well-posedness in weighted Sobolev spaces and comparison principle for the corresponding Cauchy problem   (8), (9)
are established. With sufficient smoothness of the initial data the weak solutions
 are classical ones.

In this paper we use the comparison principle for classical solutions $p(S,t), q(S,t)$ of the problem (8), (9), i.e
$ p \in C ((0,+ \infty) \times (o,T]) \cap C^{2,1}((0,+ \infty) \times (0,T)), \; \;
q \in C ((0,+ \infty) \times (0,T]), \; \; q_{t} \in ((0,+ \infty) \times (0,T))
 $.

\noindent {\bf{ Proposition 1  ([4])}}
{\it{Let  $( p_{1},q_{1} )$ and $( p_{0},q_{0} )$ be two classical solution of problem (8),(9) corresponding
 to terminal data $h=h_{1} (S)$ and $h=h_{0} (S)$ , respectively.
 }}
{\it{
\noindent  If there exists some positive constants $A$ and $\alpha$ such that $p_{i} (S,t)$ and $h_{i} (S), \; i=0,1$
 satisfy conditions (10), then

 $$
 \begin{array}{c}
 \inf (h_{1} - h_{0} ) \leq p_{1} (S,t) - p_{0} (S,t) \leq \sup ( h_{1}-h_{0}) ,
 \\
  \inf (h_{1} - h_{0} ) \leq q_{1} (S,t) - q_{0} (S,t) \leq \sup ( h_{1}-h_{0}) ,
  \end{array}
\eqno{(11)}
 $$

%

 In particular, let $h(S)$ be bounded from below (or from above) by a constant $h(S) \geq h_{\star} $
 (resp. $h(S) \leq h^{\star} $ and the pair $p(S,t),q(S,t)$ be a classical solution of the
 terminal problem (8),(9).
 Then

 $p(S,t)  \geq h_{\star}$  and $  q(S,t)\geq h_{\star}$ (respectively
 $ p(S,t)  \leq h^{\star}$  and $  q(S,t)\leq h^{\star}$).

 \noindent for any $S \in (0, + \infty)$ and any $t \in (0,T]$.
 }}


By making the substitutions $\tau= T-t$,
$u=\gamma R^0$ and $v=\gamma R^1$, the system (3)
 becomes

 \setcounter{equation}{11}

\begin{align}
\label{pde}
\begin{array}{l}
L^{p} (u,v) \equiv u_\tau-\frac{1}{2}\sigma^2S^2 u_{SS}+ ae^ue^{-v}- b= 0, \\
L^{0} (u,v) \equiv v_\tau+ ce^ve^{-u}-c=0,
\end{array}
\end{align}
where $a=\nu_{01}, b = d_0+\nu_{01}, c= \nu_{10}$. In accordance with (9) we take the initial conditions to be
\begin{equation}
 u(0,S)=u_0(S)=\gamma h(S),\quad v(0,S)=v_0(S)=\gamma h(S).
\end{equation}
For a call option,

\begin{equation}
h(S)=\max(S-K,0) .
\end{equation}


We assume ground conditions for $u,v$ of the form (10).
In the next sections, the analysis of the difference approximations
of problem(12)-(14)
will use the following comparison principle that follows from those one for $(p,q)$:

\noindent   {\bf{Proposition 2}}
{\it{Let $(\overline{u},\overline{v}),(\underline{u},\underline{v})\in C(\left[0,T\right)\times \left(0,+\infty\right) )\cap C^{2,1}
(\left(0,T\right) \times \left(0,+\infty\right)  )$ be two pairs of classical solutions of (12)-(14)
 corresponding to the initial data $h=\overline{h}$ and $h=\underline{h}$, respectively and such that conditions
 of the type (10) hold. If the following inequalities also hold:

 \setcounter{equation}{14}

\begin{equation}
 L^{p} (\overline{u}, \overline{v} ) \geq L^{p} (\underline{u}, \underline{v} ),\; L^{0} (\overline{u}, \overline{v} ) \geq L^{0} (\underline{u}, \underline{v} ) \mbox{ and } \overline{h}\geq \underline{h},
\end{equation}
then $$\overline{u}\geq \underline{u},\overline{v}\geq \underline{v} . $$

}}

Hereinbelow we will use the following canonical form of writing a 3-point  difference scheme

$$
\begin{array}{c}
A_{i}y_{i-1} - C_{i}y_{i} + B_{i}y_{i+1} = - F_{i}, \; \;
i= 1,2, \dots, N-1  \\
y_{0} = \mu_{1} , \; \; y_{N} = \mu_{2} .
\end{array}
\eqno{(16)}
$$

The discrete comparison principle for problem (16) was proved in [6,7] and is
formulated in the following way.

%
%
%
%
%

\begin{lemma}
Let the conditions

$$
A_{i}>0, \; \; B_{i}>0, \; \; D_{i}=C_{i} - A_{i} - B_{i} \geq 0, \; \;
i=1,2, \dots , N-1
\eqno{(17)}
$$

\noindent be fulfilled. Then the solution of the difference scheme (15) satisfies the inequalities

$$
y_{i} \geq 0 , \; \;
i=0 , \dots, N
 , \; \; \mbox{if} \; \; F_{i} \geq 0 , \; \;
i=1, \dots ,N-1
, \; \; \mu_{1} \geq 0, \; \; \mu_{2} \geq 0;
$$

$$
y_{i} \leq 0 , \; \;
i = 0, \dots , N
 , \; \; \mbox{if} \; \; F_{i} \leq 0 , \; \;
i = 1 , \dots , N-1
, \; \; \mu_{1} \leq 0, \; \; \mu_{2} \leq 0 .
$$

\end{lemma}

\begin{lemma}
 Let the conditions

$$\vert A_{i} \vert  \geq 0, \; \; \vert B_{i} \vert \geq 0, \; \; D_{i}=\vert C_{i} \vert  - \vert  A_{i} \vert  - \vert B_{i} \vert  > 0, \; \;
i=1, \dots , N-1
$$


\noindent be met. Then for the solution the problem (16) the estimate holds


$$
\Vert y \Vert_{C (   \overline { \Omega }^{N}   )} \leq
\max
\left\{
\vert \mu_{1} \vert ,
\vert \mu_{2} \vert ,
\left\Vert
\frac{F}{D}
\right\Vert_{C ( \Omega^{N}  )}
\right\} .
$$

%

%
%
%
%
%
%
%
%
%
%
%
%
%

\end{lemma}
\section{Implicit-Explicit Linear Scheme}
In this section, we develop a linear IMEX scheme to solve the coupled semi-linear parabolic-ordinary system problem (11)-(12).

%

\noindent For call option one possible pair of boundary conditions is, see e.g. [10,11]

\setcounter{equation}{17}

\begin{equation}
 u(\tau,0)
 = \varphi_{l} (\tau)
 =0,\quad u(\tau,S)
 = \varphi_{r} (\tau)
 \approx S_{\max}\; \mbox{ for large }S.
\end{equation}
The left natural boundary condition for $u$ is
\begin{equation}
 u_\tau(\tau,0)= -ae^{-(v(\tau,0)-u(\tau,0))}+b.
\end{equation}
On $Q_{T} = \Omega \times [0,T]$ we introduce the uniform
mesh  $ w_{ S \tau } = w_{S} \times w_{ \tau } :$

$$
{\overline{w}}_{S} = \{ S_{i} = i \triangle S , \; \; \triangle S > 0, \; \;
i=0,1, \dots , I; \; \; I \triangle S =S_{max }\} , \; \;
{\overline{w}}_{S} = w_{S} \cup \{ S_{0} , S_{I} \} ;
$$

$$
{\overline{w}}_{ \tau } = \{ \tau_{j} = j \triangle \tau , \; \;
\triangle \tau > 0, \; \; j=0,1, \dots , J; \; \;
J \triangle \tau =T \}. , \; \;
{\overline{w}}_{\tau} = w_{\tau} \cup \{ \tau_{0} , \tau_{J} \}
$$

On the discrete domain $w_{S\tau} $ we approximate the problem
(12)-(14)
by the difference scheme

$$
L^{p} (U,V) =
\frac{
U_{i}^{j+1} - U_{i}^{j}
}
{\triangle\tau } -
\frac{1}{2}
\sigma^{2}
S_{i}^{2}
\frac{
U_{i-1}^{j+1} - 2U_{i}^{j+1} + U_{i+1}^{j+1}
}
{(\triangle S)^{2}  }
+ a e^{   -V_{i}^{j} }
e^{   U_{i}^{j} } - b=0 ,
\eqno{(20)}
$$

$$
i=1,2, \dots , I-1;
$$

$$
L^{0} (U,V) =
\frac{
V_{i}^{j+1} - V_{i}^{j}
}
{\triangle\tau }+
c
e^{   -U_{i}^{j} }
e^{   V_{i}^{j} } -c =0 , \; \; i=0,1, \dots , I,
\eqno{(21)}
$$

$$
j=0,1, \dots , J-1 ;
$$

$$
U_{i}^{0} = U_{0} (S_{i} ) , \; \;
i=0,1, \dots , I ;\;
\eqno{(22)}
$$

$$
U_{0}^{j} = \varphi_{l}
(\tau_j) , \; \;
U_{I}^{j} =
\varphi_{r} ( \tau_{j}) , \; \;
j =0,1, \dots , J;
\eqno{(23)}
$$

$$
V_{i}^{0} =
V_{0} (S_{i} ) , \; \;
i=1, \dots , I.
\eqno{(24)}
$$

The natural boundary condition can be approximated as follows

$$
U_{0}^{ j+1 } = U_{0}^{j} - \triangle \tau
(a e^{u}  e^{   - V_{0}^{j}  } e^{   U_{0}^{j}  } - b) .
$$

\noindent On the $(j+1)$-th,
$j=0,1, \dots ,J-1$
time level
the scheme (20)-(23) has the form

\setcounter{equation}{24}

\begin{align}
\label{}
\begin{array}{l}
- A_{i} U_{i-1}^{j+1}  +
C_{i} U_{i}^{j+1} -
B_{i} U_{i+1}^{j+1} = F_{i},\\
\\
V_{i}^{j+1}= V_{i}^{j}-
\triangle \tau
ce^{-U_{i}^{j}+V_{i}^{j}}+c,
\end{array}
\end{align}

\noindent where

$$
A_{i} =B_{i} =  \frac{1}{2}\sigma^{2}  \frac{S_{i}^{2}}{(\triangle S)^2},\;
C_{i} =
\frac{1}{\triangle \tau}
+ A_{i} + B_{i} ,
$$

$$
F_{i}=
\frac{1}{\triangle \tau}
U_{i}^{j}-a e^{ -V_{i}^{j}}e^{ U_{i}^{j} }+b, \; \;
i=1, \dots , I-1;
$$


For the truncation error corresponding to (20)
 we find


$$
Tr1 = \frac{1}{2} \triangle \tau
 \frac{\partial^{2} u }{\partial \tau^{2} }
 (\tau_{j+1} - \theta_{1}  \triangle \tau, S_{i} )
 $$


 $$
 + \triangle \tau
 \left(
 \frac{\partial u }{\partial \tau }
 ( \tau_{j+1} - \rho^{-}  \triangle \tau, S_{i-1} ) +
 \frac{\partial u }{\partial \tau }
 ( \tau_{j+1} - \rho  \triangle \tau, S_{i} ) +
 \frac{\partial u }{\partial \tau }
 ( \tau_{j+1} - \rho^{+}  \triangle \tau, S_{i+1} )
 \right)
 $$

 $$
 +
  \frac{1}{24} ( \triangle S )^{2}
  \left\{
   \frac{\partial^{4} u }{\partial S^{2} }
 (\tau_{j+1} , S_{i}+ \theta^{+}_{1}  \triangle S )
 +
 \frac{\partial^{4} u }{\partial S^{2} }
 (\tau_{j+1} , S_{i}- \theta^{-}_{1}  \triangle S )
  \right\}
  $$

  $$
  + \triangle \tau
   \frac{\partial V }{\partial \tau }
 ( \tau_{j+1} - {\widetilde{\rho}}  \triangle \tau, S_{i} )
 e^{ u( \tau_{j+1} , S_{i} )    }
  =
 \frac{\partial u }{\partial \tau }
 ( \tau_{j+1} -
 {\stackrel{\approx}{\rho}}
    \triangle \tau, S_{i} )
   e^{ -v( \tau_{j+1} , S_{i} )    }
  $$

  $$
  =
  O ( \triangle \tau ) + (\triangle S )^{2}  .
  $$

  For the truncation error corresponding to (21) we get

  $$
Tr2 = \frac{1}{2} \triangle \tau
 \frac{\partial^{2} u }{\partial \tau^{2} }
 (\tau_{j+1} - \theta_{2}  \triangle \tau, S_{i} )+
 \triangle \tau
 \frac{\partial u }{\partial \tau }
 ( \tau_{j+1} - {\widetilde{\eta}}  \triangle \tau, S_{i} )
 e^{ v( \tau_{j+1} , S_{i} )    }
 $$

 $$
  -
 \frac{\partial v }{\partial \tau }
 ( \tau_{j+1} -
 {\stackrel{\approx}{\eta}}
    \triangle \tau, S_{i} )
   e^{ -u( \tau_{j+1} , S_{i} )    }
 = O ( \triangle \tau ).
 $$

  $$
  0 < \rho , \; \rho^{-}, \rho^{+},
  {\widetilde{ \rho}} ,  {\stackrel{\approx}{\rho}}
   <1 , \;
  0 < \theta_{1} ,
  \theta^{-}_{1} ,
  \theta^{+}_{1} < 1.
  $$


In accordance with
Notation
we define the strong norms on the meshes  ${\overline{w}}_{S} $ and $w_{S\tau} $, respectively,

$$
\Vert z \Vert_{  C ( {\overline{w}}_{S}  )  } =
\max_{  0 \leq i \leq I }
\vert z_{i} \vert , \; \;
\Vert z \Vert_{  C ( {\overline{w}}_{S \tau }   )  } =
\max_{
 { 0 \leq i \leq I }
  \atop
 { 0 \leq i  \leq J }
  }
\vert z_{i}^{j}  \vert .
$$

Let denote



$$
C_{u} = \sup_{  (\tau,S) \in Q_{T}  }
\vert u (\tau , S) \vert , \; \;
C_{v} = \sup_{  (\tau,S) \in Q_{T}  }
\vert v (\tau , S) \vert .
$$

 \noindent {\bf {Theorem 1}}
{\it{Suppose that there exists classical solution $(u,v) \in C^{2.4} (Q_{T})$ of problem (10)-(14).
%
Then for sufficiently small
$ \triangle S $ and $ \triangle \tau $ the following
error estimate holds:

$$
\Vert u -U \Vert_{   C (  w_{  S \tau }  )  } +
\Vert v -V \Vert_{   C (  w_{  S \tau }  )  }
\leq C ( \triangle \tau + ( \triangle S )^{2} ),
\eqno{(26)}
$$

\noindent where the constant $C$ doesn't depend of
$ \triangle S $ and $ \triangle \tau $.
}}


{\bf{ Proof   }}
Define  the errors $\varepsilon_{i}^{j} , \mu_{i}^{j} $  by

$$
\varepsilon_{i}^{j} = U_{i}^{j} - u ( \tau_{j} , S_{i} ) , \; \;
\mu_{i}^{j} = V_{i}^{j} - v ( \tau_{j} , S_{i} ) , \; \;
i=1, \dots , I .
$$

\noindent Then $ \{ \varepsilon_{i}^{j} \} , \{  \mu_{i}^{j} \} $ satisfy the linear system of  algebraic
equations:

$$
A_{i}
\varepsilon_{i-1}^{j+1} - C_{i}
\varepsilon_{i}^{j+1} +
B_{i}
\varepsilon_{i+1}^{j+1}
 =
F_{i} , \; \;
i=1 , \dots , I-1,
$$

$$
\varepsilon_{0}^{j+1}=0, \; \;
\varepsilon_{I}^{j+1}=0,
$$

\noindent where

$$
F_{i}^{j}=\frac{1} {  \triangle \tau   }
\varepsilon_{i}^{j} + \alpha_{i}^{j}
$$

\noindent and

$$
\mu_{i}^{j+1}= \mu_{i}^{j}
+ \triangle \tau \beta_{i}^{j} .
$$

\noindent Here $ \alpha_{i}^{j} $  and $ \beta_{i}^{j} $are the local truncation errors corresponding to the difference
equations (20) and (21), respectively. They will be estimated as follows.

Let us derive the truncation error corresponding to nonlinear (right) part:

For the nonlinear, right hand side of the first equation we obtain


%

  $$
  e^{ - v_{i}^{j}} e^{  u_{i}^{j}} =
  e^{ - \mu_{i}^{j}    - v (   \tau_{j}, S_{i} ) }  e^{ \varepsilon_{i}^{j} +  u  ( \tau_{j}, S_{i} )}
  $$

  $$
   =
  ( 1 - \mu_{i}^{j} + O (( \mu_{i}^{j} )^{2}  ) ( 1 + \varepsilon_{i}^{j} +
  O (( \varepsilon_{i}^{j} )^{2} )
  e^{ - v ( S_{i} , \tau_{j} )  }
  e^{u ( S_{i} , \tau_{j} )  }
  $$

  $$
  = ( 1 + \varepsilon_{i}^{j} - \mu_{i}^{j} )
  e^{ - v ( \tau_{j}, S_{i} )  }
  e^{u ( \tau_{j}, S_{i} )  }
  -
  O(\varepsilon_{i}^{j} \mu_{i}^{j} )+
  O (( \varepsilon_{i}^{j} )^{2} )
  + O (( \mu_{i}^{j} )^{2} ) ,
  $$

\noindent and another one form

 $$
 e^{ -V_{i}^{j} }
 e^{ U_{i}^{j} }
 = e^{ - v ( \tau_{j}, S_{i} )  }
  e^{u ( \tau_{j}, S_{i} )  }+
  O ( \varepsilon_{i}^{j}) + O ( \mu_{i}^{j} ).
  $$


Now, taking into account the Tr1, we have

  $$\alpha_{i}^{j} = O ( \triangle \tau ) +  ( \triangle S)^{2} +
  ( \varepsilon_{i}^{j} - \mu_{i}^{j} )
  e^{ - v ( \tau_{j}, S_{i} )  }
  e^{u ( \tau_{j}, S_{i} )  }
  + O ( \varepsilon_{i}^{j} \mu_{i}^{j})  + O (( \varepsilon_{i}^{j})^{2} ) + O (( \mu_{i}^{j})^{2} ).  $$

\noindent In a similar way, we find

$$
\beta_{i}^{j} =
O ( \triangle \tau ) +
( \mu_{i}^{j} - \varepsilon_{i}^{j} )
e^{
- u(   \tau_{j}, S_{i} )
}
e^{
v(    \tau_{j}, S_{i} )
}
+ O ( \varepsilon_{i}^{j} \mu_{i}^{j} )+
O (( \varepsilon_{i}^{j})^{2} ) +
O (( \mu_{i}^{j})^{2} ).
$$

   Applying Lemma 2.1 we get

   $$
   \Vert \varepsilon_{i}^{j+1} \Vert_{C} \leq \triangle \tau
   \Vert {\widehat{F}} \Vert ,
   $$

\noindent where $\Vert \cdot \Vert_{C}$ is the strong norms $C (   {\overline{w}}_{S}  )$ as defined above.

   \noindent We estimate $ \Vert F^{j+1} \Vert: $

   $$
   \Vert F^{j+1} \Vert \leq  ( \frac{1} {  \triangle \tau } + e^{ C_{u} C_{v}}  )
   \Vert \varepsilon^{j} \Vert  + e^{ C_{u} C_{v}}
   \Vert \mu^{j} \Vert
   $$

   $$
   + O ( \triangle \tau ) + O ( ( \triangle S )^{2} ) +
   O ( \Vert \varepsilon^{j} \Vert^{2} )+
   O ( \Vert \mu^{j} \Vert^{2} )).
   $$

   \noindent Next,

   $$
   \Vert \mu^{j+1} \Vert \leq  \Vert \mu^{j} \Vert +
   \triangle \tau
    ( e^{ C_{u} C_{v}}
   \Vert \varepsilon^{j} \Vert  +
    O ( \triangle \tau ) +
   O ( \Vert \varepsilon^{j} \Vert^{2} )+
   O ( \Vert \mu^{j} \Vert^{2} )).
   $$

   \noindent Therefore,

   $$
   \Vert \varepsilon^{j+1} \Vert +
   \Vert \mu^{j+1} \Vert \leq
   ( 1  + 2 \triangle \tau e^{C_{u}} e^{ C_{v}}  )
   \Vert \varepsilon^{j} \Vert
   $$

   $$
   + ( 1  +  \triangle \tau C_{u} C_{v}  )
   \Vert \mu^{j} \Vert + \triangle \tau
    ( O ( \triangle \tau ) +  ( \triangle S )^{2}  +
   O ( \Vert \varepsilon^{j} \Vert^{2}  )+
   O ( \Vert \mu^{j} \Vert^{2}  )).
   $$

\noindent For $j=0$ we have
  $ \varepsilon_{i}^{0}=0 , \; \mu_{i}^{0}=0 $ and
  then

  $$ \alpha_{i}^{0} = O ( \triangle \tau ) + (\triangle S)^{2}, \; \; \beta_{i}^{0} = O (\triangle \tau ) .$$

   Since $  \Vert \varepsilon^{0} \Vert = \Vert \mu^{0} \Vert =0 $ , we get

   $$
   \Vert \varepsilon^{1} \Vert = C \triangle \tau  ( \triangle  \tau + ( \triangle S)^{2} ) , \; \;
   \Vert \varepsilon^{1} \Vert = C \triangle \tau  ( \triangle  \tau + ( \triangle S)^{2} ) .
   $$

   \noindent Therefore, by induction we have

   $$
   \Vert \varepsilon^{j+ 1} \Vert
   + \Vert \mu^{j+ 1} \Vert  \leq
   ( 1  + 2 \triangle \tau C_{u} C_{v}  )
   ( \Vert \varepsilon^{j} \Vert +
   \Vert \mu^{j} \Vert )
   $$

   $$
      + \triangle \tau C ( \triangle  \tau + ( \triangle S)^{2} ).
      $$

      \noindent which implies that

       $$
   \Vert \varepsilon^{j+1} \Vert +
   \Vert \mu^{j+1} \Vert \leq
   C \sum_{k=0}^{j}
   ( 1  + 2 \triangle \tau C_{u} C_{v}  )^{k}  \triangle \tau
   (  \triangle \tau + ( \triangle S)^{2} )
   $$

   $$
   C
   ( \triangle \tau + ( \triangle S)^{2} )
   ((
   ( 1  + 2 \triangle \tau C_{u} C_{v}  )^{J}
   ( 1  + 2 \triangle \tau C_{u} C_{v}  )^{-1}
   $$

   $$
  \leq  C( \triangle \tau + ( \triangle S)^{2} ) \hfill \Box
   $$

The following discrete comparison principle for the $(U,V)$ is crucial for the positivity of
the discrete approximations of the indeference prices $p$ and $q$ on the base of the scheme (20)-(24).

\noindent {\bf{Theorem 2}}
{\it{Let the assumptions of  Theorem 1 hold
Let also $ (  { \overline {U}} , { \overline {V}} ) $ , $ (  { \underline {U}} , { \underline {V}} )$
be grid functions defined on
${\overline{
  w}}_{S \tau}
  $
   and the inequalities hold:

$$
L^{p}  (  { \overline {U}} , { \overline {V}} )
\geq L^{p}
(  { \underline {U}} , { \underline {V}} ) , \; \;
L^{0}  (  { \overline {U}} , { \overline {V}} )
\geq L^{0}
(  { \underline {U}} , { \underline {V}} ) ,
\eqno{(17)}
 $$

 $$
    { \overline {U}}_{i}^{0}  \geq  { \underline {U}}_{i}^{0}  , \; \;
  { \overline {V}}_{i}^{0}  \geq  { \underline {V}}_{i}^{0}
  , \; \;
  i=0 , \dots , I,
\eqno{(28)}
 $$

$$
    { \overline {V}}_{0}^{j}  \geq  { \underline {V}}_{0}^{i}  , \; \;
  { \overline {U}}_{M}^{j}  \geq  { \underline {U}}_{M}^{j}
  , \; \;
  j=1 , \dots , J .
\eqno{(29)}
 $$

 \noindent Then for sufficiently small $ \triangle S  $ and $ \triangle \tau  $ we have

 $$
    { \overline {U}}_{i}^{j}  \geq  { \underline {U}}_{i}^{j}  , \; \;
  { \overline {V}}_{i}^{j}  \geq  { \underline {V}}_{i}^{j}
  , \; \;
  i=0 ,1 \dots , I , \; \;
  j=0,1 , \dots , J .
\eqno{(30)}
 $$
 }}

 {\bf {Proof.     }} Let introduce

 $$
 y_{i}^{j} =
 { \overline {U}}_{i}^{j}  - { \underline {U}}_{i}^{j} ,
 \; \;
  z_{i}^{j} =
 { \overline {V}}_{i}^{j}  -  { \underline {V}}_{i}^{j} , \; \;
   \; \;
  i=0 ,1 \dots , I , \; \;
  j=0,1 , \dots , J.
  $$

  Then, from (26) we obtain

  $$
\frac{
y_{i}^{j+1} - y_{i}^{j}}{ \triangle \tau } -\frac{1}{2}\sigma^{2}S_{i}^{2}\frac{y_{i-1}^{j+1} -2y_{i}^{j+1} + y_{i+1}^{j+1}}{(\triangle S)^{2}}+ a( e^{ -  {\overline{V}}_{i}^{j} }e^{ {\overline{U}}_{i}^{j} } - e^{ -  {\underline{V}}_{i}^{j} }e^{   {\underline{U}}_{i}^{j} } )\geq 0 ,
\eqno{(31)}
$$

$$
\frac{z_{i}^{j+1} - z_{i}^{j}}{ \triangle \tau }+c(e^{-{\overline{U}}_{i}^{j} }e^{ {\overline{V}}_{i}^{j} }- e^{ -  {\underline{U}}_{i}^{j} }e^{   {\underline{V}}_{i}^{j} } )\geq 0, \; \;i=1 \dots , I-1 , \; \;  j=1 , \dots , J-1,
\eqno{(32)}
$$
Using the mean-value theorem we get
\begin{align}
 e^{ -  {\overline{V}}_{i}^{j} }e^{   {\overline{U}}_{i}^{j} } - e^{ -  {\underline{V}}_{i}^{j} }
e^{   {\underline{U}}_{i}^{j} } =e^{ -  {\widetilde{V}}_{i}^{j} }e^{   {\widetilde{U}}_{i}^{j} }
( y_{i}^{j} - z_{i}^{j}), \nonumber\\
{\widetilde{U}}_{i}^{j} ={\underline{U}}_{i}^{j} +{\widetilde{\theta }}({\overline{U}}_{i}^{j} -
{\underline{U}}_{i}^{j}),\; {\widetilde{V}}_{i}^{j} = {\overline{V}}_{i}^{j} +{\widetilde{\theta }}(
{\overline{V}}_{i}^{j} -
{\underline{V}}_{i}^{j}),\; 0 <{\widetilde{\theta }} < 1 , \nonumber\\
e^{ -  {\overline{U}}_{i}^{j} }e^{   {\overline{V}}_{i}^{j} } - e^{ -{\underline{U}}_{i}^{j} }
e^{   {\underline{V}}_{i}^{j} } =e^{ -  {\widehat{U}}_{i}^{j} }e^{   {\widehat{V}}_{i}^{j} }
( z_{i}^{j} - y_{i}^{j}), \nonumber\\
{\widehat{U}}_{i}^{j} ={\underline{U}}_{i}^{j} +{\widehat{\theta }}({\overline{U}}_{i}^{j} -
{\underline{U}}_{i}^{j}),\;
{\widehat{V}}_{i}^{j} =
{\overline{V}}_{i}^{j} +
{\widehat{\theta }}
(
{\overline{V}}_{i}^{j} -
{\underline{V}}_{i}^{j}), \; \;
0 <
{\widehat{\theta }} < 1 . \nonumber
\end{align}

We rewrite (30) in the form

$$
 A_{i} y_{i-1}^{j+1}  -
C_{i} y_{i}^{j+1}
B_{i} y_{i+1}^{j+1} \geq - F_{i} ,
$$
$$
A_{i} = \frac{1}{2}\sigma^{2}S_{i}^{2}\frac{\triangle \tau}{\triangle S},\;
B_{i} = \frac{1}{2}\sigma^{2}S_{i}^{2}\frac{\triangle \tau}{\triangle S},\;
C_{i} =\frac{1} {   \triangle \tau }+ A_{i} + B_{i} ,
$$

$$
F_{i}=
\frac{1} {   \triangle \tau }
y_{i}^{j}-
ae^{ -  {\widetilde{V}}_{i}^{j} }e^{   {\widetilde{U}}_{i}^{j} }
( y_{i}^{j} - z_{i}^{j}).
$$

Next, we rewrite (31) in the form

$$
\frac{
z_{i}^{j+1}
}
{  \triangle \tau  }
\geq
\left(\frac{1}{\triangle \tau}-ce^{ -  {\widehat{U}}_{i}^{j} }e^{   {\widehat{V}}_{i}^{j} }  \right) z_{i}^{j} +
ce^{ -  {\widehat{U}}_{i}^{j} }e^{   {\widehat{V}}_{i}^{j} }y_{i}^{j}.
\eqno{(33)}
$$

We apply the method of mathematical induction with respect to $j$
to prove that

$$
y_{i}^{j} \geq 0, \; \; z_{i}^{j} \geq 0, \; \; i =0,1, \dots , I , \; \;
j=0,1, \dots , J .
\eqno{(34)}
$$

From (16), (17), we have

$$
y_{i}^{o} \geq 0, \; \; z_{i}^{o} \geq 0, \; \; i =0,1, \dots , I , \; \;
$$

\noindent Assuming that (32) holds when $j=k-1$, we will show that for $j=k$
the above inequalities are
  true.

On the base of Theorem 1 we can confirm that
for sufficiently small $\triangle \tau , \triangle S $
there exists constants $C_{u} , C_{v}$, such that

$$
\max ( \Vert {\overline{U}} \Vert , \Vert {\underline{U}} ) \leq 2
C_{u}, \; \;
\max ( \Vert {\overline{V}} \Vert , \Vert {\underline{V}} ) \leq 2 C_{v}.
$$

\noindent Then, if it necessary, we choose
$ \triangle \tau $ in additional smaller such that

$$\triangle \tau < \min (a,c) e^{   2 C_{u}   }
e^{   2 C_{v}   }
\eqno{(35)}
$$

By induction, $ y_{i}^{k-1} \geq 0 , \; z_{i}^{k-1} \geq 0 $ and using  (Lemma 2.1)
we conclude that $F_{i} \geq 0 , \; i=0, 1, \dots , I-1 $. Now Lemma 2.1
implies $y_{i}^{k} \geq 0 , \; i=0,1 , \dots , I$.  It is clear from (32) and  (34) that
$ z_{i}^{k} \geq 0 , \; i=1, \dots , I-1.  \hfill \Box $

\section{Implicit-Explicit Linearised Scheme}
Let us consider first the implicit scheme:
$$
\frac{
U_{i}^{j+1} - U_{i}^{j}
}
{\triangle\tau } -
\frac{1}{2}
\sigma^{2}
S_{i}^2
\frac{
U_{i-1}^{j+1} - 2U_{i}^{j+1} + U_{i+1}^{j+1}
}
{(\triangle S)^{2}  }
+ a e^{   -V_{i}^{j+1} }
e^{   U_{i}^{j+1} } - b=0,
\eqno{(36)}
$$
$$
\frac{
V_{i}^{j+1} - V_{i}^{j}
}
{\triangle\tau }+
c
e^{   -U_{i}^{j+1} }
e^{   V_{i}^{j+1} } -c =0 , \; \; \eqno{(37)}
$$$$
i=1,2, \dots , I-1; \; \;
j=0,1, \dots , J-1 .
$$

\noindent with boundary and initial approximations (22)-(24).

\noindent By Taylor expansion we get 

$$
e^{U_{i}^{j+1}-V_{i}^{j+1} }=e^{-V_{i}^{j} }e^{U_{i}^{j}}(1+V_{i}^{j}-U_{i}^{j})+e^{-V_{i}^{j} }e^{U_{i}^{j}}(U_{i}^{j+1}-V_{i}^{j+1})
$$

$$
+O((U_{i}^{j+1}-U_{i}^{j})^2)+O((V_{i}^{j+1}-V_{i}^{j})^2),
$$

$$
e^{V_{i}^{j+1}-U_{i}^{j+1} }=e^{-U_{i}^{j} }e^{V_{i}^{j}}(1-V_{i}^{j}+U_{i}^{j})+e^{-U_{i}^{j} }e^{V_{i}^{j}}(V_{i}^{j+1}-U_{i}^{j+1})
$$

$$
+O((U_{i}^{j+1}-U_{i}^{j})^2)+O((V_{i}^{j+1}-V_{i}^{j})^2).
$$
We drop the $O$-terms
 and the  results we insert in (36) and (37) to obtain:

\setcounter{equation}{37}

\begin{align}
- &      \hat{A}_{i} U_{i-1}^{j+1} +\hat{C}_{i} U_{i}^{j+1}-\hat{B}_{i} U_{i+1}^{j+1}+\hat{D}_{i} V_{i}^{j+1} = \hat{F}_{i},\\
\nonumber \\
  &       \hat{E}_i U_{i}^{j+1}+\hat{K}_iV_{i}^{j+1}= G_{i} ,
\end{align}

\noindent where \begin{align}
\hat{A}_{i} &= B_{i}= \frac{1}{2}\sigma^{2}
\frac{S_{i}^{2} }
{(\triangle S)^2},\;
\hat{C}_{i} =
\frac{1}{  \triangle \tau } + A_{i} + B_{i}
+a  e^{U_{i}^{j}-V_{i}^{j}},\nonumber\\
\hat{D}_{i} &= -a \tau e^{U_{i}^{j}-V_{i}^{j}},\;
\hat{F}_{i}=
\frac{1}{\triangle \tau}
U_{i}^{j}-a \tau e^{U_{i}^{j}-V_{i}^{j}}(1+V_{i}^{j}-U_{i}^{j})+b\triangle \tau , \nonumber\\
\hat{E}_{i} &=-c\triangle \tau e^{V_{i}^{j}-U_{i}^{j}},\;\hat{K}_{i} =
\frac{1}{\triangle \tau}
+c e^{V_{i}^{j}-U_{i}^{j}} , \nonumber\\
G_{i}&=
\frac{1}{\triangle \tau}
V_{i}^{j}-c \tau e^{V_{i}^{j}-U_{i}^{j}}(1-V_{i}^{j}+U_{i}^{j})+c .  \nonumber
\end{align}

Since $ a e^{   U_{i}^{j} - V_{i}^{j}    } > 0$ , the diagonal domination can significally increase in
comparison with IMEX linear scheme, see system (20),(21).

\noindent {\bf{Theorem 3}}
{\it{Let the assumptions of Theorem 1 hold.Then
suppose that there exists classical solution $(u,v) \in C^{2.4} (Q_{T})$ of problem (10).
%
Then for sufficiently small
$ \triangle S $ and $ \triangle \tau $ the following
error estimate holds:

$$
\Vert u -U \Vert_{   C (  w_{  S \tau }  )  } +
\Vert v -V \Vert_{   C (  w_{  S \tau }  )  }
\leq C ( \triangle \tau + ( \triangle S )^{2} ),
$$

\noindent where the constant $C$ doesn't depend of
$ \triangle S $ and $ \triangle \tau $.
}}


{\bf{ Proof.  }}
Substituting $V_{i}^{j+1}$ from (37)
 into (36) the first one  we get

$$
\begin{array}{l}
- \hat{A}_{i} U_{i-1}^{j+1} +\left(\hat{C}_{i}-\dfrac{\hat{D}_{i}\hat{E}_{i}}{\hat{K}_{i}}\right) U_{i}^{j+1}-\hat{B}_{i} U_{i+1}^{j+1}= \hat{F}_{i}-\dfrac{\hat{D}_{i}}{\hat{K}_{i}}\tilde{F}_{i} ,\\
V_{i}^{j+1}= \dfrac{G_{i}}{\hat{K}_{i}}-\dfrac{\hat{E}_{i}}{\hat{K}_{i}}U_{i}^{j+1} , \; \; i=1 , \dots , I-1
\end{array}
$$

\noindent with $U_{i}^{0} , \; i=0,1, \dots , I$ and $ U_{0}^{j}, \;  U_{I}^{j}  , \; j=0,1, \dots , J $ given by (22),(23) and (24).

For the error we have the linear system of algebraic equations

$$
-A_{i} \varepsilon_{i-1}^{j+1} +
\left(
C_{i} -
\frac{D_{i}E_{i}  }{ K_{i} }
\right)
\varepsilon_{i}^{j+1} - B_{i}
\varepsilon_{i+1}^{j+1} =
{\widehat{F}}_{i+1} =
\frac{
\varepsilon_{i}^{j}
 }
 {  \triangle \tau }
 +
 \alpha_{i}^{j} ,
 $$

 $$
 \varepsilon_{0}^{j+1}=0 , \; \;
 \varepsilon_{I}^{j+1}=0
 $$

 $$
 \mu_{i}^{j+1} =
 - \frac{ E_{i} }{  K_{i} }
 \varepsilon_{i}^{j+1} +
 \frac{ \mu_{i}^{j} }{  \triangle \tau } +
 \beta_{i}^{j} , \; \;
 i=1, \dots , I-1.
 $$

 Further we follow the line of Theorem 1 to complete the proof.  $ \hfill \Box$

 The scheme (36), (37) also has similar
 comparison properties of the linear IMEX scheme described in Theorem 2.


\section{Numerical Experiments}

In the section we perform numerical experiments to illustrate the accuracy, effectiveness and convergence of the implicit-explicit
linear scheme (20)-(24) (Scheme 1) and implicit-explicit linearized scheme (38),(39) (Scheme 2) developed in this article. We
provide experiments both with uniform and non-uniform meshes. Also, we present results of numerical experiments using Richardson
extrapolation in time.

The Tables (presented results) show the
accuracy in maximal discrete norm $\Vert \cdot \Vert$ and convergence rate at final time $T$, using
{\it{two consecutive meshes with formulas}}

$$
\mbox{Ratio} = \log_{2}
 (E_{ I/2 }^{w}   /  E_{I}), \; \;
E_{I}^{w} = \Vert w_{ex} - W \Vert_{ \infty} ,
$$

\noindent where $w_{ex}$ and $W$ are the exact and
the corresponding numerical solutions, respectively. In our case  $w_{ex}$ is $R^{0}$ or $R^{1}$ .

In Tables \ref{table:one}, \ref{table:two},
we give the results from the computations
IMEX linear Scheme 1.
\begin{table}[ht]
\caption{Convergence results for at the money ($S = 2, K= 2, S_{\min}=0$ and $S_{\max} = 5$) and
$\triangle \tau = \triangle S /2$
based on Scheme 1}
\centering
\begin{tabular}{c| c c c |c c c} 
\hline 
&\multicolumn{3}{c|}{$R^0$}&\multicolumn{3}{c}{$R^1$}\\
$I$ & Value & Difference & Ratio & Value & Difference & Ratio \\
\hline
30  & 0.246669 &         &            & 0.235165 &       & \\
60  & 0.247438 & 7.70e-04&            & 0.235917 &7.52e-04 & \\
120 & 0.247749 & 3.11e-04& 2.48 (1.31)& 0.236218 &3.01e-04 & 2.50 (1.32)\\
240 & 0.247887 & 1.38e-04& 2.25 (1.17)& 0.236349 &1.31e-04 & 2.30 (1.20)\\
480 & 0.247952 & 6.50e-05& 2.12 (1.09)& 0.236410 &6.10e-05 & 2.15 (1.10)\\
960 & 0.247983 & 3.10e-05& 2.10 (1.07)& 0.236439 &2.90e-05 & 2.10 (1.07)\\
\hline
\end{tabular}
\label{table:one}
\end{table}


\begin{table}[ht]
\caption{Convergence results for at the money ($S = 2, K= 2, S_{\min}= 0$ and $S_{\max} = 5$) and taking $ \triangle \tau = \triangle S_{i}  /2$ and using nonuniform Tavella-Randal grid with $\alpha=15$ based on Scheme 1}
\centering
\begin{tabular}{c| c c c |c c c} 
\hline 
&\multicolumn{3}{c|}{$R^0$}&\multicolumn{3}{c}{$R^1$}\\
$I$ & Value & Difference & Ratio & Value & Difference & Ratio \\
\hline
30  & 0.247196 &         &             & 0.235660 &       & \\
60  & 0.247863 & 6.67e-04&             & 0.236305 &6.45e-04 & \\
120 & 0.248124 & 2.61e-04& 2.56 (1.35) & 0.236552 &2.47e-04 & 2.61 (1.38)\\
240 & 0.248238 & 1.14e-04& 2.29 (1.20) & 0.236657 &1.05e-04 & 2.35 (1.23)\\
480 & 0.248291 & 5.30e-05& 2.15 (1.10) & 0.236706 &4.90e-05 & 2.14 (1.10)\\
960 & 0.248322 & 3.10e-05& 1.71 (0.77) & 0.236735 &2.90e-05 & 1.69 (0.76)\\
\hline
\end{tabular}
\label{table:two}
\end{table}
Table 2
is based on a non-uniform grid and also shows that the scheme is first order in time. Here we use Tavella-Randal [8]  mesh:
\begin{align}
 S_i &= K+\alpha\left(c_2\frac{i}{I}+c_1\left(1-\frac{i}{I}\right)\right),\nonumber\\
c_1 &=\sinh^{-1}\left(\frac{S_{\min}-K}{\alpha}\right), \; \;
c_2
=\sinh^{-1}\left(\frac{S_{\max}-K}{\alpha}\right). \nonumber
\end{align}
In this case, we choose to concentrate mesh points around the strike price $K$ since we expect the error to be largest there.
In Table 3 we list the results from computation with Scheme 2
  that for this non-uniform grid the results are still first order accurate in time as in the uniform case.

\begin{table}[ht]
\caption{Convergence results for at the money ($S = 2, K= 2, S_{\min}=0$ and $S_{\max} = 5$) and taking $\triangle \tau = \triangle S/2$ based on Scheme 2}
\centering
\begin{tabular}{c| c c c |c c c} 
\hline 
&\multicolumn{3}{c|}{$R^0$}&\multicolumn{3}{c}{$R^1$}\\
$I$ & Value & Difference & Ratio & Value & Difference & Ratio \\
\hline
30  & 0.246685 &         &            & 0.234952 &       & \\
60  & 0.247444 & 7.59e-04&            & 0.235812 &8.60e-04 & \\
120 & 0.247752 & 3.08e-04& 2.46 (1.30)& 0.236165 &3.53e-04 & 2.44 (1.28)\\
240 & 0.247889 & 1.37e-04& 2.25 (1.17)& 0.236323 &1.58e-04 & 2.23 (1.16)\\
480 & 0.247953 & 6.40e-05& 2.14 (1.10)& 0.236397 &7.40e-05 & 2.14 (1.09)\\
960 & 0.247984 & 3.10e-05& 2.06 (1.05)& 0.236433 &3.60e-05 & 2.06 (1.04)\\
\hline
\end{tabular}
\label{table:five}
\end{table}

\begin{table}[ht]
\caption{Convergence results for at the money ($S = 2, K= 2, S_{\min}=0$ and $S_{\max} = 5$) and taking $\triangle \tau =\triangle S_{i} /2$ and using nonuniform Tavella-Randal grid with $\alpha=15$ based on scheme 2}
\centering
\begin{tabular}{c| c c c |c c c} 
\hline 
&\multicolumn{3}{c|}{$R^0$}&\multicolumn{3}{c}{$R^1$}\\
$I$ & Value & Difference & Ratio & Value & Difference & Ratio \\
\hline
30  & 0.248722  &             &             & 0.237005  &       & \\
60  & 0.249432  & 7.10e-04    &             & 0.237812  &8.07e-04 & \\
120 & 0.249715  & 2.83e-04    & 2.51 (1.33) & 0.238139  &3.27e-04 & 2.47 (1.30)\\
240 & 0.249839  & 1.24e-04    & 2.28 (1.19) & 0.238283  &1.44e-04 & 2.27 (1.18)\\
480 & 0.249897  & 5.80e-05    & 2.14 (1.10) & 0.238351  &6.80e-05 & 2.12 (1.08)\\
960 & 0.249928  & 3.10e-05    & 1.87 (0.90) & 0.238387  &3.60e-05 & 1.89 (0.92)\\
\hline
\end{tabular}
\label{table:six}
\end{table}

Now, we improve the convergence in time applying
Richardson extrapolation [4].
 To this aim
 we use the formula $$Y_n=\frac{2^pW_n-Z_n}{2^p-1}$$ where $p$ is order of numerical solution (1 in our case) and $W_n$ is the solution obtained using time step
 $\triangle \tau/2$ and $Z_n$ is the solution obtained using time step $\triangle \tau$. The resulting solution $Y_n$ has order of accuracy $p+1$ 
 [4].
  Table \ref{table:nine} shows the result of applying this technique to the {\it{Scheme 1.}} The order of accuracy in time is now two.  Similarly this technique is applied to Scheme 2, see  Table 6.
 Hence the convergence is much slower {\it{but smoother}} compared to the explicit based {\it{Scheme 1}} due the
 {\it{error of linearisation.}}
  The tables shows  second order in time.
\begin{table}[ht]
\caption{Convergence results for at the money ($S = 2, K= 2, S_{\min}=0$ and $S_{\max} = 5$) and taking $\triangle \tau =\triangle S /2$ based on Scheme 1 using Richardson extrapolation
}
\centering
\begin{tabular}{c c c c c c} 
\hline 
$I$ &$Z_n,W_n$ & $Y_n$ & Difference & Ratio (order)\\
\hline
10 & 0.2451080 &&&\\
20 & 0.2465578 & 0.2480075 &&\\
40 & 0.2472811 & 0.2480045 & 3.02e-6 &&\\
80 & 0.2476431 & 0.2480051 & 5.79e-7 & 5.22 (2.38)\\
160 & 0.2478242 & 0.2480053 & 1.96e-7 & 2.96 (1.56)\\
320 & 0.2479148 & 0.2480053 & 5.13e-8 & 3.82 (1.93)\\
640 & 0.2479600 & 0.2480053 & 1.27e-8 & 4.05 (2.02)\\
1280 & 0.2479827 & 0.2480053 & 3.08e-9 & 4.12 (2.04)\\
2560 & 0.2479940 & 0.2480053 & 7.45e-10 & 4.13 (2.05)\\
\hline
\end{tabular}
\label{table:nine}
\end{table}

\newpage

\begin{table}[ht]
\caption{Convergence results for at the money ($S = 2, K= 2, S_{\min}=0$ and $S_{\max} = 5$) and taking $\triangle \tau =\triangle S/2$ based on Scheme 2 using Richardson extrapolation.
}
\centering
\begin{tabular}{c c c c c c} 
\hline 
$I$ &$Z_n,W_n$ & $Y_n$ & Difference & Ratio (order)\\
\hline
10&0.2451717  &&&\\
20& 0.2465832 & 0.2479947 &&\\
40&0.2472928 & 0.2480023 & 7.64e-6 &\\
80&0.2476486 & 0.2480045 & 2.14e-6 & 3.57 (1.84) \\
160&0.2478269 & 0.2480051 & 6.22e-7 & 3.44 (1.78)\\
320&0.2479161 & 0.2480053 & 1.78e-7 & 3.49 (1.81)\\
640&0.2479607 & 0.2480053 & 4.93e-8 & 3.62 (1.85)\\
1280&0.2479830 & 0.2480053 & 1.32e-8 & 3.73 (1.90)\\
2560& 0.2479942 & 0.2480053 & 3.46e-9 & 3.81 (1.93)\\
5120&0.2479998 & 0.2480053 & 8.95e-10 & 3.87 (1.95)\\
10240&0.2480026 & 0.2480053 & 2.29e-10 & 3.91 (1.97)\\
\hline
\end{tabular}
\label{table:ten}
\end{table}


\begin{figure}[ht]
\centering
{\small{
\subfigure[$p$ at $t=0$ and $t=T$]{%
\includegraphics[height=.2\textheight]{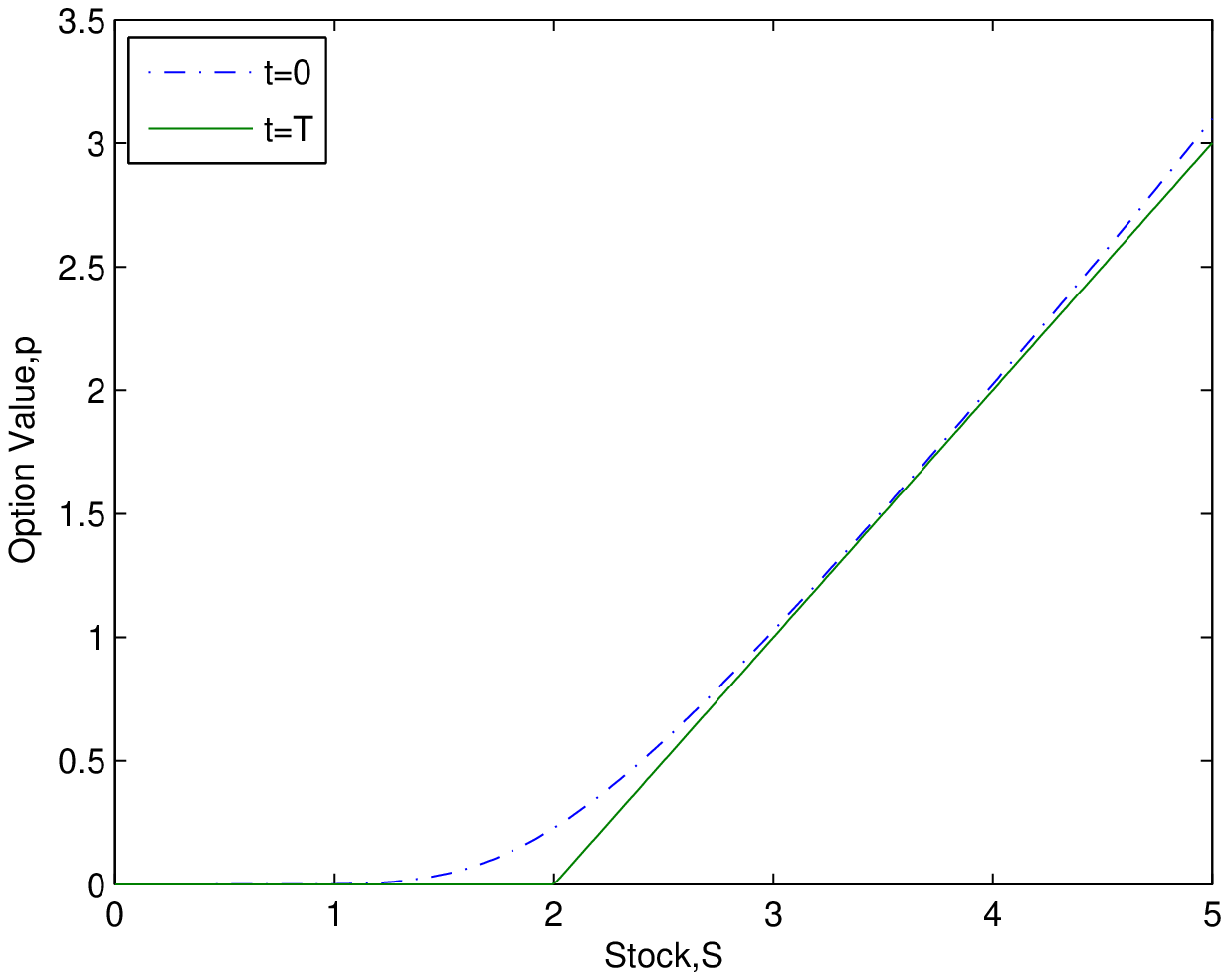}
\label{fig:subfigure1}}
\;
\subfigure[$q$ at $t=0$ and $t=T$]{%
\includegraphics[height=.2\textheight]{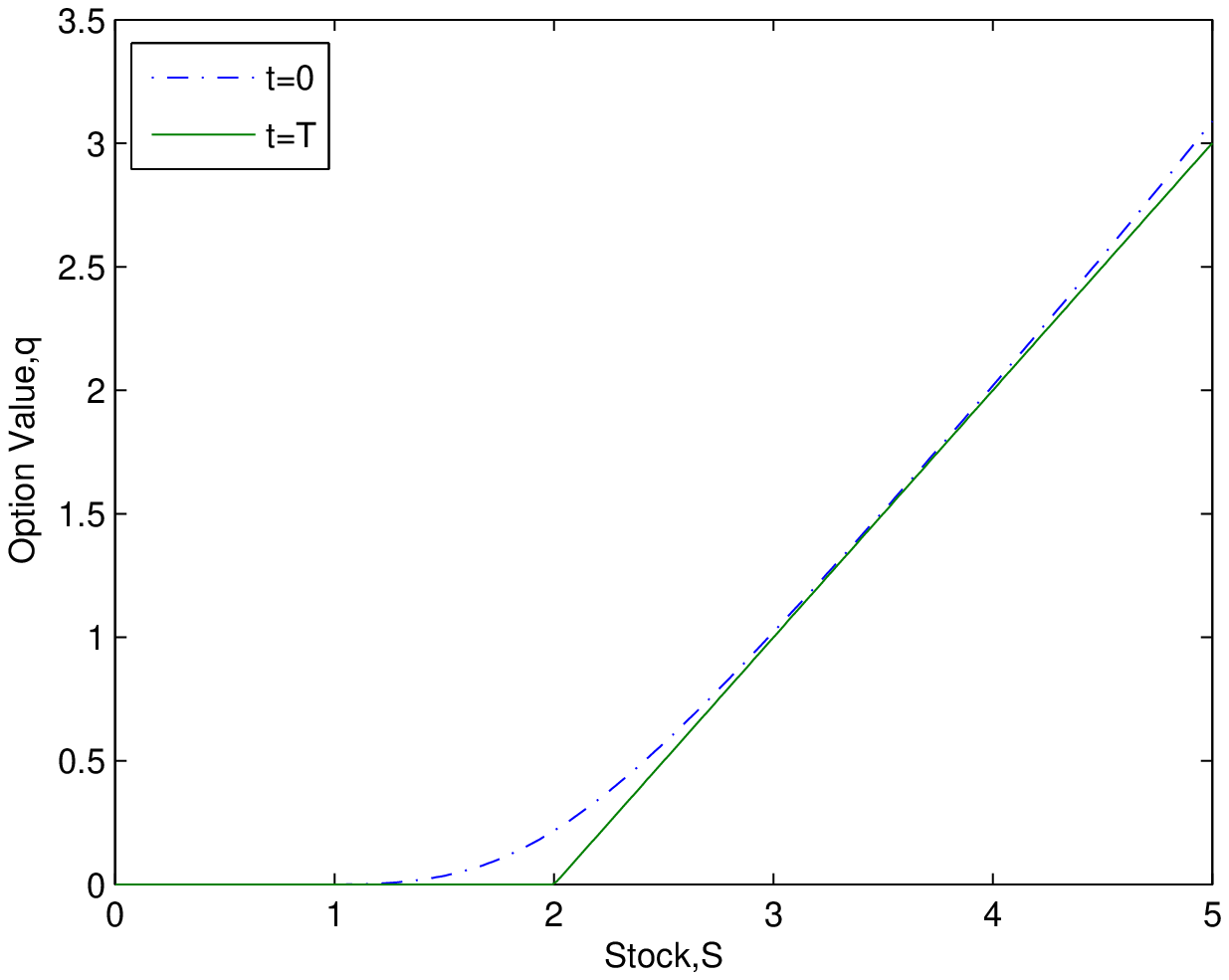}
\label{fig:subfigure2}}
\caption{Comparing European option values at issue and maturity in the liquid and illiquid states for the IMEX Linear scheme}
\label{fig:maturity1}
}}
\end{figure}
\begin{figure}[ht]
\centering
{\small{
\subfigure[$p$ at $t=0$ and $t=T$]{%
\includegraphics[height=.2\textheight]{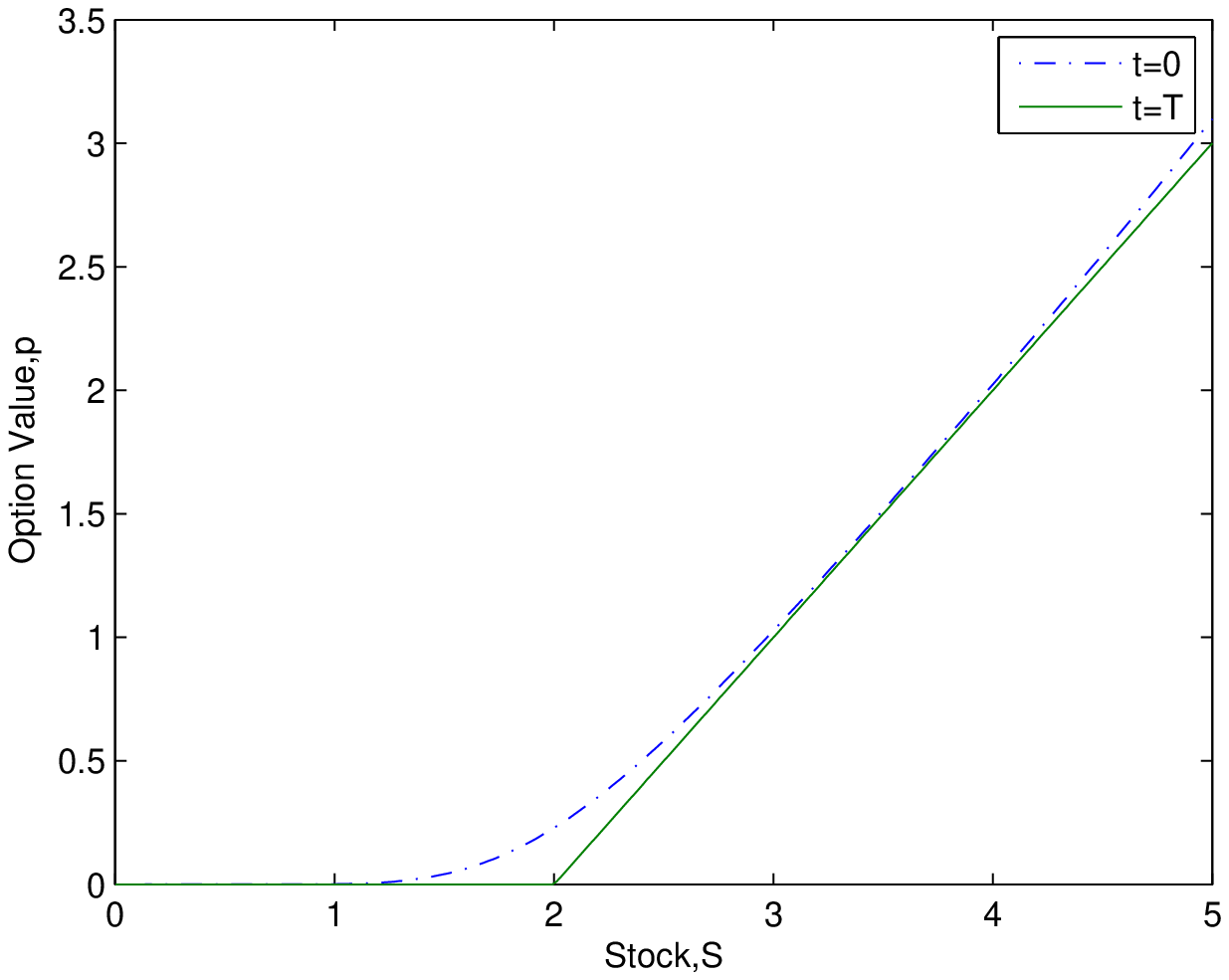}
\label{fig:subfigure1}}
\;
\subfigure[$q$ at $t=0$ and $t=T$]{%
\includegraphics[height=.2\textheight]{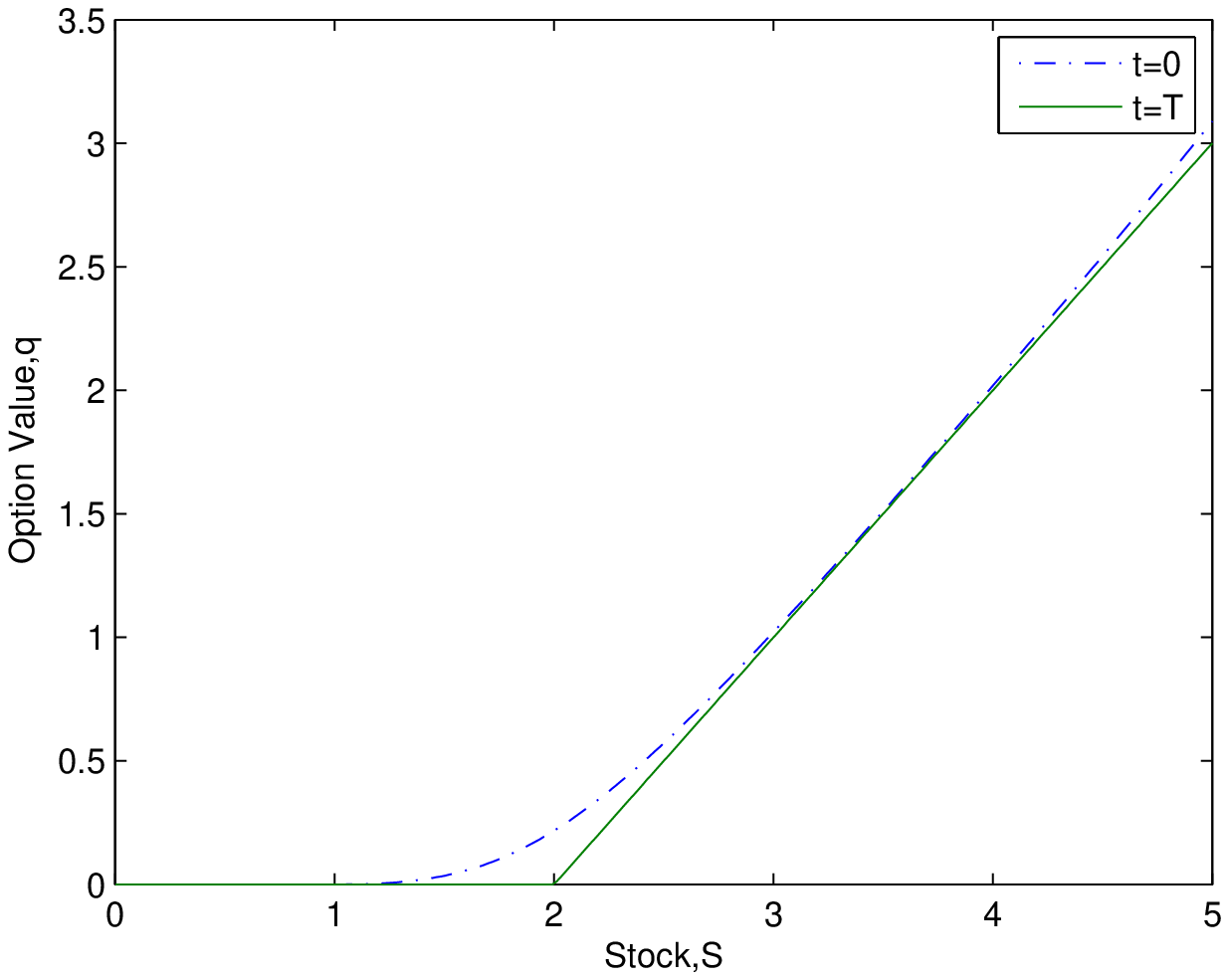}
\label{fig:subfigure2}}
\caption{Comparing European option values at issue and maturity in the liquid and illiquid statesfor the IMEX Linearised scheme}
\label{fig:maturity2}
}}
\end{figure}
In Figure \ref{fig:maturity1} we compare options values $p$ and $q$ at issue and maturity in the liquid and illiquid states using the parameters $\mu=0.06$,  $\sigma$=0.3, $\nu_{01}=1$, $\nu_{10}=12$, $K=2$, $T=1$, $S_{\max}=5$ and $\gamma=1$ using the
Scheme 1
.
Figure \ref{fig:maturity2}
illustrate
 the linearised scheme,  using the same parameters.
 Figures 1,2
 illustrate
  the positivity of the solution $(p, q)$, using both schemes.


\section{Conclusions}

{\small{
In this work we have considered  one-dimensional problem of European options with liquidity shocks.
We have constructed and analyzed two IMEX finite difference schemes that preserve the positivity property
of the differential solution.
The second one(the IMEX linearized scheme) has better diagonal domination, respectively monotonicity.
It would be interesting to consider extensions of the IMEX schemes to the American options
with liquidity shocks. In this case one has to solve a free boundary problem. It could be written
as a linear complementary problem which could be discretized using the schemes given here. The
extension is beyond the scope of this paper, and we leave it for further work.

{\bf{Acknowledgement}}The authors thank to Prof.M.Koleva for the help at the numerical experiments. This research was supported
by the European Union under Grant Agreement number 304 617 (FP7 Marie Curie Action Project Multi-ITN Strike-Novel Methods in
Computational Finance) and Bulgarian National Fund of Science under Project DFNI I02/20-2014.
}}



%

\end{document}